# Antiferromagnetic Quantum Anomalous Hall Effect Modulated by Spin Flips and Flops

Zichen Lian[1†], Yongchao Wang[1†], Yongqian Wang[2,3], Yang Feng[4], Zehao Dong[1], Shuai Yang[2,3], Liangcai Xu[1], Yaoxin Li[1], Bohan Fu[2,3], Yuetan Li[1], Wanjun Jiang[1], Chang Liu[2,3*], Jinsong Zhang[1,6*], Yayu Wang[1,5,6*]

[1]*State Key Laboratory of Low Dimensional Quantum Physics, Department of Physics, Tsinghua University, Beijing 100084, P. R. China*

[2]*Beijing Key Laboratory of Opto-electronic Functional Materials & Micro-Nano Devices, Department of Physics, Renmin University of China, 100872 Beijing, China*

[3]*Key Laboratory of Quantum State Construction and Manipulation (Ministry of Education), Renmin University of China, Beijing 100872, China*

[4]*Beijing Academy of Quantum Information Sciences, Beijing 100193, P. R. China*

[5]*New Cornerstone Science Laboratory, Frontier Science Center for Quantum Information, Beijing 100084, P. R. China*

[6]*Hefei National Laboratory, Hefei 230088, China*

[†] *These authors contributed equally to this work.*

\* Emails: liuchang_phy@ruc.edu.cn; jinsongzhang@tsinghua.edu.cn;

yayuwang@tsinghua.edu.cn

**The interplay between nontrivial band topology and layered antiferromagnetism in MnBi$_2$Te$_4$ has opened up a new avenue for exploring topological phases of matter. Representative examples include the quantum anomalous Hall effect and axion insulator state observed in odd and even number layers of MnBi$_2$Te$_4$, when the top and bottom surfaces have parallel and antiparallel spin alignments respectively. The rich and complex spin dynamics associated with the van der Waals antiferromagnetic order is expected to generate novel topological phases and phase transitions that are unique to MnBi$_2$Te$_4$. Here we fabricate a device of 7-septuple-layer MnBi$_2$Te$_4$ covered with AlO$_x$ capping layer, which enables the investigation of antiferromagnetic quantum anomalous Hall effect over wide parameter spaces. By tuning the gate voltage and perpendicular magnetic field, we uncover a cascade of quantum phase transitions that can be attributed to the influence of spin configurations on charge transport. Furthermore, we find that an in-plane magnetic field enhances both the coercive field and exchange gap of the surface state, in sharp contrast to that in ferromagnetic quantum anomalous Hall state. We propose that these peculiar features arise from the spin flip and flop transitions inherent to van der Waals antiferromagnet. The versatile tunability of the quantum anomalous Hall effect in MnBi$_2$Te$_4$ paves the way for potential applications in topological antiferromagnetic spintronics.**

The quantum anomalous Hall (QAH) effect realizes one-dimensional dissipationless chiral edge state transport in topological materials in the absence of external magnetic field[1-3]. It has garnered significant attentions not only for the exotic transport properties, but also for potential applications in quantum metrology[4,5] and topological quantum computation[6-8]. Over the past decade, the pursuit of new materials exhibiting the QAH effect at higher temperature (*T*) or with novel types of quantization has been a rapidly developing field[3,9-19]. Among the various systems explored, MnBi$_2$Te$_4$ represents the only one hosting the QAH effect in the presence of bulk antiferromagnetic (AFM) order, thus can be dubbed the AFM QAH state. As shown in Fig. 1a, the Mn moments in each septuple layer (SL) have ferromagnetic (FM) alignment, but exhibit interlayer AFM order[20-28]. As has been proposed theoretically, different spin stacking sequences of the SLs and various metamagnetic phases will lead to fundamentally different bulk band structure and edge state conduction[29-32]. Recently, even-odd layer-dependent magnetism and the long-sought surface spin flop transition have been observed in MnBi$_2$Te$_4$ (Ref. [33-35]), which trigger the search for novel QAH behavior modulated by the spin dynamics. However, there has been little experimental progress along this direction, mainly owing to the challenge in obtaining high-quality device with zero-field quantization[12]. It has been shown that the MnBi$_2$Te$_4$ crystals are prone to various types of defect[36-43], and the nano-fabrication process may introduce further complications[44].

In this work, we employ a new device architecture based on 7-SL MnBi$_2$Te$_4$ with significantly improved performance. The MnBi$_2$Te$_4$ single crystal is grown by solid-state reaction, and its high Néel temperature $T_N \sim 26$ K indicates high crystalline quality. We notice that in most odd-number-SL MnBi$_2$Te$_4$ devices showing a large anomalous Hall effect (AHE), one surface of the thin film or flake is in contact with AlO$_x$, either as the exfoliation agent[12] or epitaxial substrate[45]. The AlO$_x$ layer not only provides protection against the environmental contamination and damage from the fabrication process[44], but may also stabilize the surface magnetic order by enhancing the perpendicular magnetic anisotropy like in many spintronic devices[46-48]. We implement this idea by depositing an AlO$_x$ capping layer on the exfoliated MnBi$_2$Te$_4$ flake prior to spin-coating of PMMA, as schematically sketched in Fig. 1b (see methods for details). The other side of the flake directly lies on the SiO$_2$/Si substrate, which also serves as a bottom gate dielectric.

Our device configuration enables the investigation of the AFM QAH effect over a wide range of control parameters. Figure 1c presents the gate voltage ($V_g$) dependent Hall resistivity ($\rho_{yx}$) and longitudinal resistivity ($\rho_{xx}$) measured at zero perpendicular magnetic field $\mu_0 H_z = 0$ and $T = 0.02$ K. The quantized $\rho_{yx}$ and vanishing $\rho_{xx}$ at the charge neutrality point (CNP) around $V_g = 30$ V clearly demonstrate the dissipationless nature of the chiral edge state in the QAH state. To map out the full phase diagram of the AFM QAH effect, we sweep $V_g$ at different fixed $H_z$ values. Figure 1d displays the colormap of $\rho_{xx}$ as a function of $V_g$ and $H_z$, where the blue region with small $\rho_{xx}$ represents the QAH state (see supplementary Fig. S3b for the $\rho_{yx}$ map). Interestingly, the QAH region expands suddenly as $\mu_0H_z$ increases to around 2.2 T, especially for the hole-doped side ($V_g < 30$ V) that was unexplored before. The QAH region then shows an abrupt decrease near 3.8 T, and after that broadens gradually with increasing $H_z$.

Figure 1e displays the $H_z$-dependent $\rho_{yx}$ and $\rho_{xx}$ loops at $T = 0.02$ K for representative $V_g$s, which exhibit much richer features than the FM QAH effect[3,9-11]. At $V_g = 30$ V, the sample exhibits the anticipated QAH effect with $\rho_{yx}$ nearly quantized at 0.981 $h/e^2$, and $\rho_{xx}$ dropped to 0.011 $h/e^2$ at $\mu_0H_z = 0$. At $\mu_0H_1 \sim 0.5$ T, there is a sharp plateau transition when the Chern number changes sign from $C = +1$ to -1. This is caused by the spin-flip transition from the ↓↑↓↑↓↓ (AFM1) state to the ↑↓↑↓↑↑ (AFM2) state, where ↑ and ↓ represent the magnetization of each SL. When $V_g$ deviates from the CNP, we first observe the linear behavior of $\rho_{yx}$ at low $H_z$, indicating that hole- ($V_g \leq 25$ V) and electron-type carriers ($V_g \geq 40$ V) begin to contribute to the ordinary Hall effect. With varied $V_g$ and $H_z$, a cascade of quantum phase transitions characterized by the deviation and recovery of $\rho_{yx}$ quantization start to emerge. As indicated by the arrows in the $V_g = 15$ V panel, at $\mu_0H_2 \sim 2.2$ T the $\rho_{yx}$ shows a sudden jump and returns to the $-h/e^2$ plateau, meanwhile $\rho_{xx}$ drops from a finite value to almost zero. With further increase of magnetic field to $\mu_0H_3 \sim 3.8$ T, the quantized state is suppressed into a broad regime with finite dissipation, but is recovered again above $\mu_0H_4 \sim 7.8$ T.

We then study the $T$ evolution of the $\rho_{yx}$ and $\rho_{xx}$ loops at $V_g = 30$ V, as shown in Fig. 2a (see supplementary Fig. S4 for separately displayed curves). In contrast to the quantized $\rho_{yx}$ plateau persisting over the entire $H_z$ range at $T = 0.02$ K, the high-$T$ curves exhibit much more complex $H_z$ dependence. Most of the curves exhibit four characteristic $H_z$ scales that vary with $T$, namely the $\rho_{yx}$ sign reversal transition at $H_1$, the sudden increase of $\rho_{yx}$ at $H_2$, the weakening of $\rho_{yx}$ at $H_3$, and the gradual recovery above $H_4$. They can be regarded as the finite $T$ version of the cascaded quantum phase transitions at the ground state tuned by $V_g$. To directly visualize the $T$ dependent phase transitions, we plot the variation of $\rho_{yx}$ and its derivative versus $H_z$ in the $T$–$H_z$ plane, as shown by the colored maps in Figs. 2b and 2c. Here the four white dashed lines represent the characteristic $H_z$ scales that separate the phase diagram into distinct regions, and the nature of each phase will be discussed later. To quantitatively characterize the robustness of each quantized state, we perform the Arrhenius fitting of the longitudinal conductivity ($\sigma_{xx}$) versus $T$. As shown in the inset of Fig. 2d, log($\sigma_{xx}$) exhibits a linear dependence on $1/T$, allowing us to extract the gap size by the thermal activation behavior $\sigma_{xx} \sim \exp(-\Delta/2k_BT)$, where $k_B$ is the Boltzmann constant and $\Delta$ denotes the gap size. Figure 2d shows the evolution of $\Delta$ as a function of $H_z$ at $V_g$ = 30 V. The most dramatic feature is the abrupt jump of $\Delta$ at $\mu_0H_2 \sim 2.2$ T, after which $\Delta/k_B$ increases from $\sim 4$ K to near 25 K. Above $\mu_0H_3 \sim 3.8$ T, it begins to decrease, reaches a minimum, and increases again up to $\sim 35$ K at $\mu_0H_4 \sim 7.8$ T when all moments are aligned along the $H_z$ direction. The slow suppression of $\Delta$ at even higher $H_z$ can be attributed to the opposite sign of the internal exchange field in MnBi$_2$Te$_4$ to the external $H_z$, as revealed in previous experiment[12]. Figure 2f and its inset display the $H_z$ dependence of $\rho_{yx}$ at $T$ = 2 K, which reveals a one-to-one correspondence between the $\rho_{yx}$ transitions and $\Delta$ variation. Figure 2e depicts the mapping of $\Delta$ with $H_z$ under different $V_g$s, which exhibits similar pattern as the $\rho_{xx}$ map in Fig. 1d.

Before we proceed to the next sets of experiment, it is instructive to lay down the magnetic order for each state revealed by the above results. In the QAH effect, the Hall quantization is determined by the exchange gap $\Delta = JM_zS_z$ between local moments and Dirac fermions, where $J$ represents the coupling constant, $M_z$ and $S_z$ represent the average $z$-component of local moments and itinerant electron spins, respectively. Theoretically, the $\Delta$ value in MnBi$_2$Te$_4$ is determined mainly by the magnetization of the surface SLs, and to certain extent by the sub-surface SLs as well[37]. Therefore, the evolution of $\Delta$ with $H_z$ reflects the magnetization orientations of the few surface and subsurface layers. Combined with prior magnetic measurements, we propose the schematic magnetic configurations as shown in Fig. 2g. At $H_1$, the spin-flip process corresponds to a reversal of spins in all SLs, resulting in the switch of Chern number between ±1 with $\Delta$ intact. Above $H_4$, the moments in all SLs are polarized into the FM state, leading to a robust Chern insulator phase with large $\Delta$. While these two limits are well-understood, the metamagnetic phases in the intermediate regimes are most intriguing. The $\mu_0H_3 \sim 3.8$ T field scale coincides with the bulk spin flop transition that has been demonstrated by polar reflective magnetic circular dichroism (RMCD)[33,35], after which all SLs enter the canted AFM (cAFM) state that can exhibit the QAH effect in thin flakes[26]. During this process, the z-component of bulk magnetization increases with $H_z$ but is reduced at the surfaces, leading to decreased gap size (ref.[33]). The $\mu_0 H_2 \sim 2.2$ T transition is the most puzzling one, which causes a dramatic increase of $\Delta$. We notice that such field scale is very close to the surface spin flop transition in 6-SL MnBi$_2$Te$_4$ with bare top surface[33,35,49], which should be absent in odd-SL devices. We propose that the top SL has a much stronger bonding with the AlO$_x$ capping layer, thus has much weaker AFM coupling with the sub-surface layer[50]. At $H_2$, the bottom 6-SL block undergoes a surface spin flop transition, similar to that revealed by RMCD measurement and theoretical calculation[33]. After that, the second layer is rotated to nearly parallel to the top layer, and the bottom layer keeps an out-of-plane magnetization. Such a magnetic configuration is highly beneficial for the QAH effect due to the enhancement of near-surface magnetization[30,31], leading to the significant increase of $\Delta$. The surface spin flop transition at $H_2$ also connects naturally to the bulk spin flop transition at $H_3$, upon which the bottom layer is tilted and causes a slight decrease of $\Delta$.

We then use another tuning knob, the in-plane magnetic field $H_y$, to manipulate the magnetic configurations in MnBi$_2$Te$_4$. Figures 3a to 3c display the $T$ evolution of $\rho_{yx}$ versus $H_z$ loops in selected $H_y$ magnitudes (the complete dataset is shown in supplementary Fig. S5). At low $T$s, the loops expand significantly with increasing $H_y$, indicating the enhancement of coercivity. The hardening of magnetism by in-plane field is in sharp contrast to that observed in the FM QAH systems[51,52]. More intriguingly, the hardening of magnetism becomes weaker and exhibits an opposite trend at higher $T$, when the coercive loop at 22 K is completely suppressed by a sufficiently large $H_y$. The hardening effect can be better visualized by directly comparing the hysteresis in varied $H_y$ at $T = 0.02$ K, as shown in Fig. 3d. In Figs. 3e and 3f, we plot the variation of $H_1$ as a function of $H_y$ at different $T$s, which clearly demonstrate the two opposite trends in different $T$ regions. The hysteresis hardening by transverse magnetic field is very rare, and the reversal of trend with $T$ makes it even more perplexing. In supplementary Fig. S11 we switch the in-plane field to the $H_x$ direction, which shows qualitatively the same behavior.

The influence of in-plane magnetic field on the layered AFM order is a highly sophisticated issue and remains largely elusive, thus we adopt a phenomenological approach to explain the $H_y$-enhanced coercivity. The discrete jumps at $H_2$ strongly indicate that the magnetization reversal is related to the domain wall displacement[53], which is sensitive to the pinning by defects. There is ample evidence that MnBi$_2$Te$_4$ crystals are prone to various types of defect[38-43], and recent magnetic imaging experiment reveals the critical role of defects in domain wall pinning[54]. As schematically drawn in Fig. 3g, in the absence of $H_y$ the defect moments are randomly orientated, thus can be regarded as weak point-like pinning centers[55]. The application of $H_y$ polarizes the defect moments along the film plane, which can form large clusters with size comparable to the domain wall width. The system enters the strong pinning regime[55], giving rise to a hardening of the hysteresis for the AFM domains in MnBi$_2$Te$_4$. In fact, it has been demonstrated that for a disordered magnet, a transverse field can enhance the domain pinning effect and harden the magnet through site-random fields[52,56]. Although the experiment was performed on disordered FM materials, the theoretical model was originally designed for AFM, hence should be applicable to MnBi$_2$Te$_4$. At higher $T$ close to $T_N$, the overall AFM order is weakened and the thermal fluctuations completely overcome the domain wall pinning barrier. The system enters the classical regime that is described by the Stoner-Wohlfarth model[55], in which the in-plane field softens the hysteresis.

In addition to the hardening of hysteresis in low $H_z$ regime, the in-plane field has more direct consequence on the AFM QAH effect at higher field. Figures 4a and 4b display the influence of $H_y$ on the $\rho_{yx}$ versus $H_z$ loops for two regimes in the phase diagram: the hole-doped region ($V_g$ = 25 V) at $T$ = 0.02 K, and the CNP region ($V_g$ = 30 V) at $T$ = 0.7 K. In the absence of $H_y$, both cases have weakened QAH effect with non-quantized $\rho_{yx}$ between $H_1$ and $H_2$. Remarkably, we find that $H_y$ not only hardens the magnetism by increasing $H_1$, but also enhances the AFM QAH effect close to $H_2$. As $H_y$ increases, the transition near $H_2$ becomes smoother and $\rho_{yx}$ becomes closer to the quantized value. Figures 4c and 4d display the phase diagrams constructed by plotting $\sigma_{xx}$ in the $V_g$ and $H_z$ plane for $\mu_0 H_y$ = 0 and 1.5

T, respectively. Notably, the application of $H_y$ enlarges the blue area with dissipationless chiral edge state transport. Figure 4e presents the evolution of $\Delta$ in different $\mu_0 H_y$ values (see supplementary Fig. S10 for the raw data), which clearly demonstrate the smoothing of the transition near $\mu_0 H_2 \sim 2.2$ T and the increase of $\Delta$ with $H_y$.

The increase of $\Delta$ under $H_y$ is consistent with our simulation based on a modified AFM spin chain model[33,35,50,57,58], as described in Supplementary Session I. For an A-type AFM with strong surface perpendicular anisotropy and reduced coupling with the subsurface layer, the application of $H_y$ tilts the down moments towards the in-plane direction, while the up moments mostly retain the z-component. Consequently, the total magnetization along the *z*-axis is enhanced by the in-plane magnetic field, as summarized in Fig. 4f, facilitating the gap opening at the Dirac point[30,31]. Additionally, the application of an in-plane magnetic field also drives the defect moments to lie in the film plane, which mitigate the magnetic disorder fluctuations and expand the effective exchange gap[37].

The 7-SL $MnBi_2Te_4$ device with an $AlO_x$ capping layer exhibits a cascade of phase transitions that can be attributed to the influences of spin flips and flops on the topological transport properties that are unique to the van der Waals type layered AFM order. The close correlation between magnetism and band topology also allows us to probe the magnetic order through edge state transport measurements. Moreover, our experiments provide important new clues for enhancing the AFM QAH effect by controlling the interfacial properties and external field orientations. Contrary to the usual expectation, an in-plane magnetic field not only hardens the hysteresis, but also enlarges the energy gap on the topological surface state. Such a powerful *in situ* knob is crucial for exploring novel topological phenomena based on the AFM QAH effect and the application of $MnBi_2Te_4$ in quantized topological AFM spintronics[59-61].

## Figure Captions

**Fig. 1 | Crystal structure, device configuration, and $V_g$ dependent transport of 7-SL MnBi$_2$Te$_4$. a,** Schematic crystal structure of MnBi$_2$Te$_4$. The red/blue rectangle represent the magnetization of up/down SLs. The red/blue arrows indicate the magnetic moment directions of Mn ions. **b,** Schematic device structure of an exfoliated 7-SL MnBi$_2$Te$_4$ flake on SiO$_2$/Si substrate and covered with the AlO$_x$ capping layer. **c,** $V_g$ dependent $\rho_{yx}$ and $\rho_{xx}$ at $\mu_0 H_z = 0$ T and $T = 0.02$ K, revealing the existence of QAH effect around $V_g = 30$ V. **d,** Color map of $\rho_{xx}$ as a function of $V_g$ and $H_z$. The blue region with diminishing $\rho_{xx}$ represents the QAH state. **e,** The $H_z$ dependence of $\rho_{xx}$ and $\rho_{yx}$ at various $V_g$s at $T = 0.02$ K. The arrows in the $V_g = 15$ V panel indicates the four characteristic field scales of the cascaded quantum phase transitions.

**Fig. 2 | Temperature dependence of $\rho_{yx}$ and $\rho_{xx}$ at $V_g = 30$ V. a,** The $H_z$-dependent $\rho_{yx}$ and $\rho_{xx}$ loops at selected temperatures at $V_g = 30$ V. **b, c,** Color maps of $\rho_{yx}$ and $\partial \rho_{yx}/\partial H_z$. The phase diagram is separated into the AFM1, AFM2, Surface Spin Flop (SSF), Canted AFM (cAFM) and FM regions by four characteristic field scales ($H_1$, $H_2$, $H_3$, $H_4$) marked by white dashed lines. **d,** The exchange gap size $\Delta$ as a function of $H_z$. The inset shows the Arrhenius plots of $\sigma_{xx}$ at $\mu_0 H_z = 0$, 3, and 9 T. **e,** Color map of $\Delta$ as a function of $H_z$ and $V_g$, which shows similar patterns to Fig. 1d. **f,** The evolution of $\rho_{yx}$ with $H_z$ at $T = 2$ K, which reveals a one-to-one correspondence between the $\rho_{yx}$ transitions and changes of $\Delta$. **g,** Schematic structures of the local moments for 7-SL MnBi$_2$Te$_4$ in varied $H_z$ derived from the simulations.

**Fig. 3 | The manipulation of magnetic hysteresis by an in-plane magnetic field. a-c,** Temperature evolution of the $\rho_{yx}$ versus $H_z$ loops with in-plane magnetic field $\mu_0 H_y = 0$ T **(a)**, 1 T **(b)** and 2.5 T **(c)**. **d,** At $T = 0.02$ K, the coercivity of the $\rho_{yx}$ versus $H_z$ loops becomes larger with increasing $H_y$. **e,f,** The coercivity as a function of $H_y$ for various temperatures. The coercive field is defined as the point where $\rho_{yx}$ crosses zero. At low $T$, the magnetism is hardened by $H_y$, but at high $T$ it shows an opposite trend. **g,** Schematic of enhanced domain wall pinning by $H_y$. The defect moments are driven to lie in the plane by $H_y$, forming larger clusters and stronger pinning centers for the AFM domain walls.

**Fig. 4 | The enhancement of QAH state by an in-plane magnetic field.** The $H_z$ dependent $\rho_{yx}$ at $V_g = 25$ V and $T = 0.02$ K **(a)**, $V_g = 30$ V and $T = 0.7$ K **(b)** under different $H_y$. In both cases the QAH state between $H_1$ and $H_2$ is enhanced by in-plane magnetic field. **c, d,** The color map of $\sigma_{xx}$ as a function of $H_z$ and $V_g$ with $\mu_0 H_y = 0$ and 1.5 T, respectively, at $T = 0.02$ K. The blue region characteristic of the QAH state is widened by $H_y$. **e,** The evolution of $\Delta$ with $H_z$ under different $H_y$ amplitudes, which reveals the increase of gap size by the in-plane magnetic field. **f,** The simulated total magnetization along the z-axis under different $H_y$ based on a modified AFM spin chain model. It shows similar behavior as $\Delta$ in **e**, indicating the enhancement of QAH effect by in-plane magnetic field.

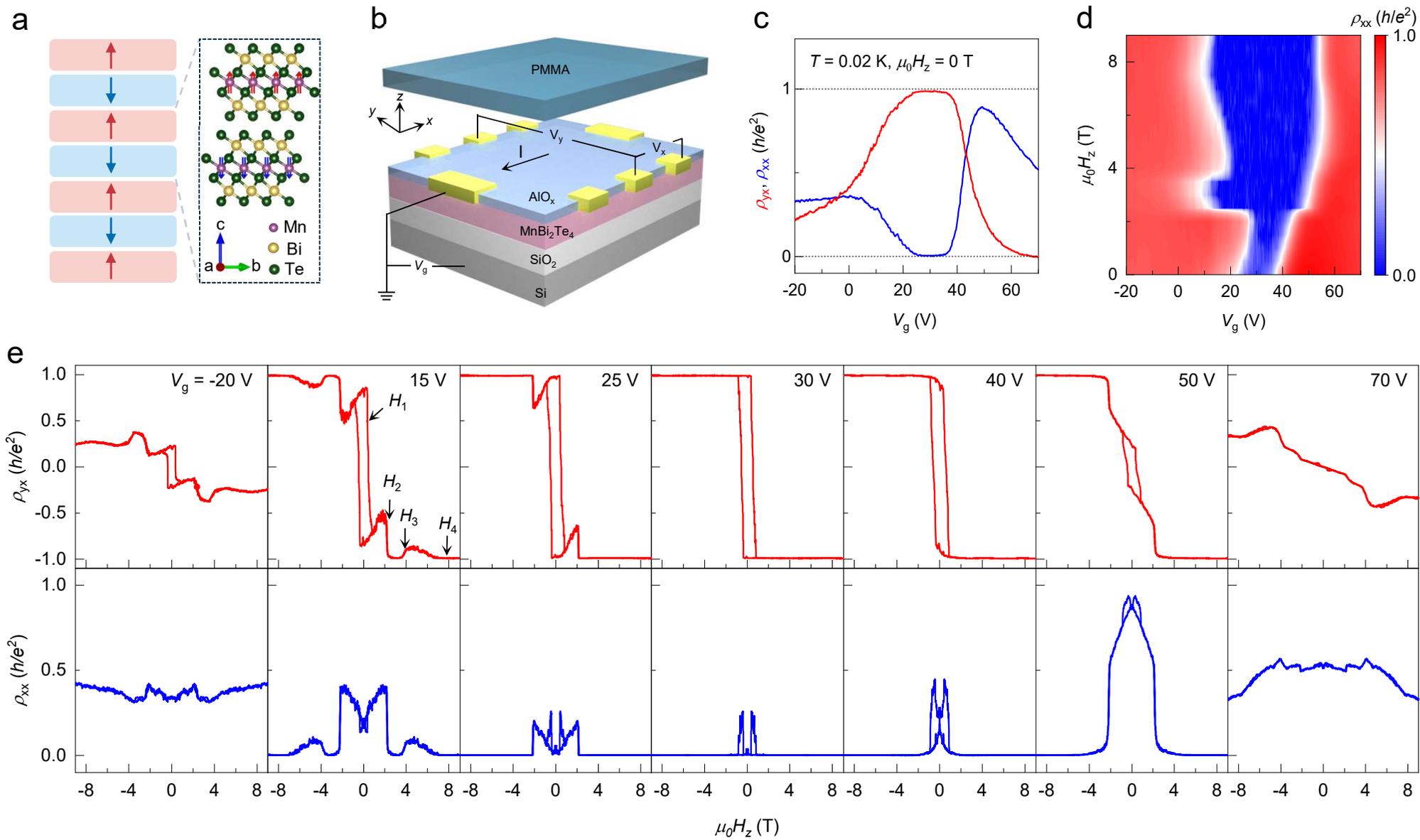

Fig. 1

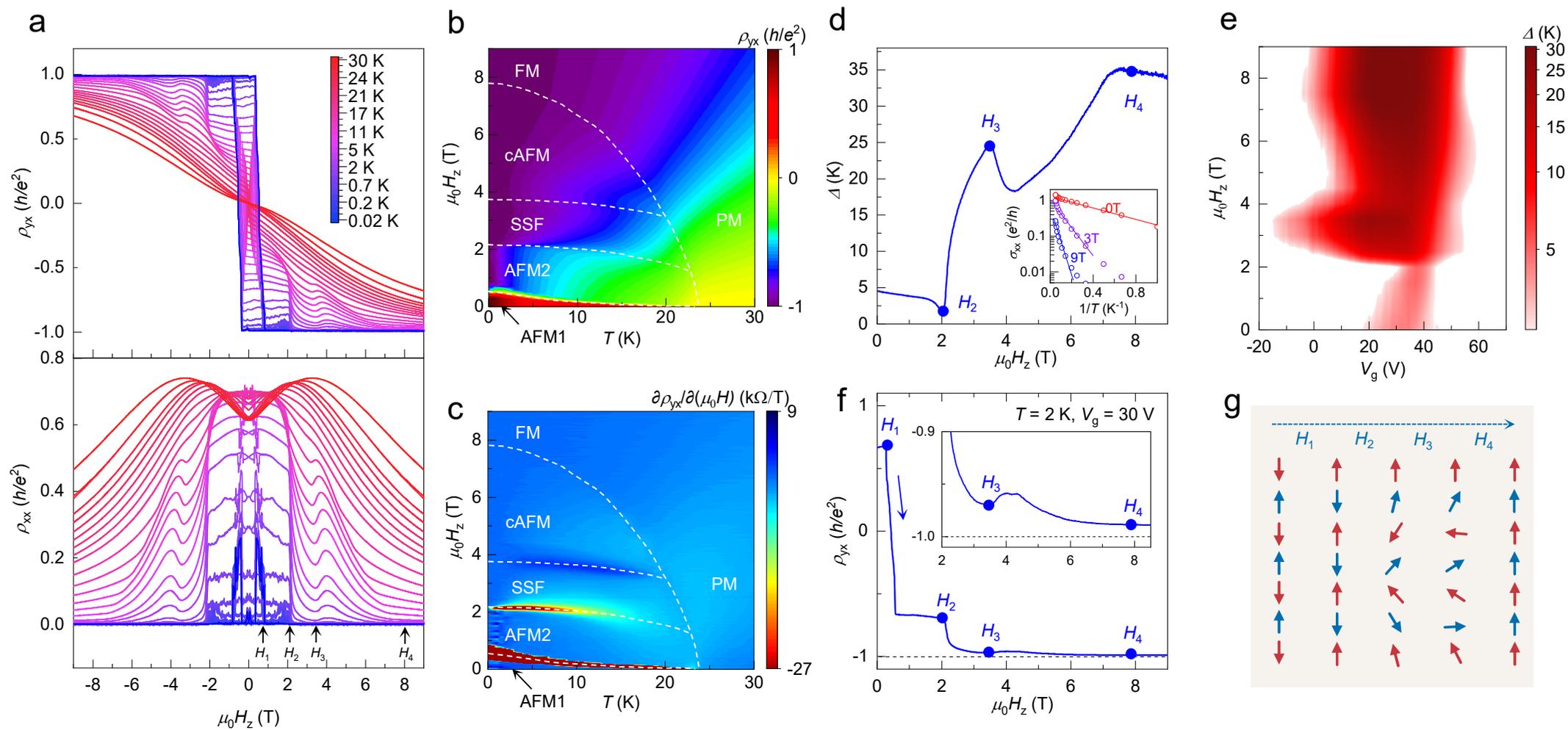

Fig. 2

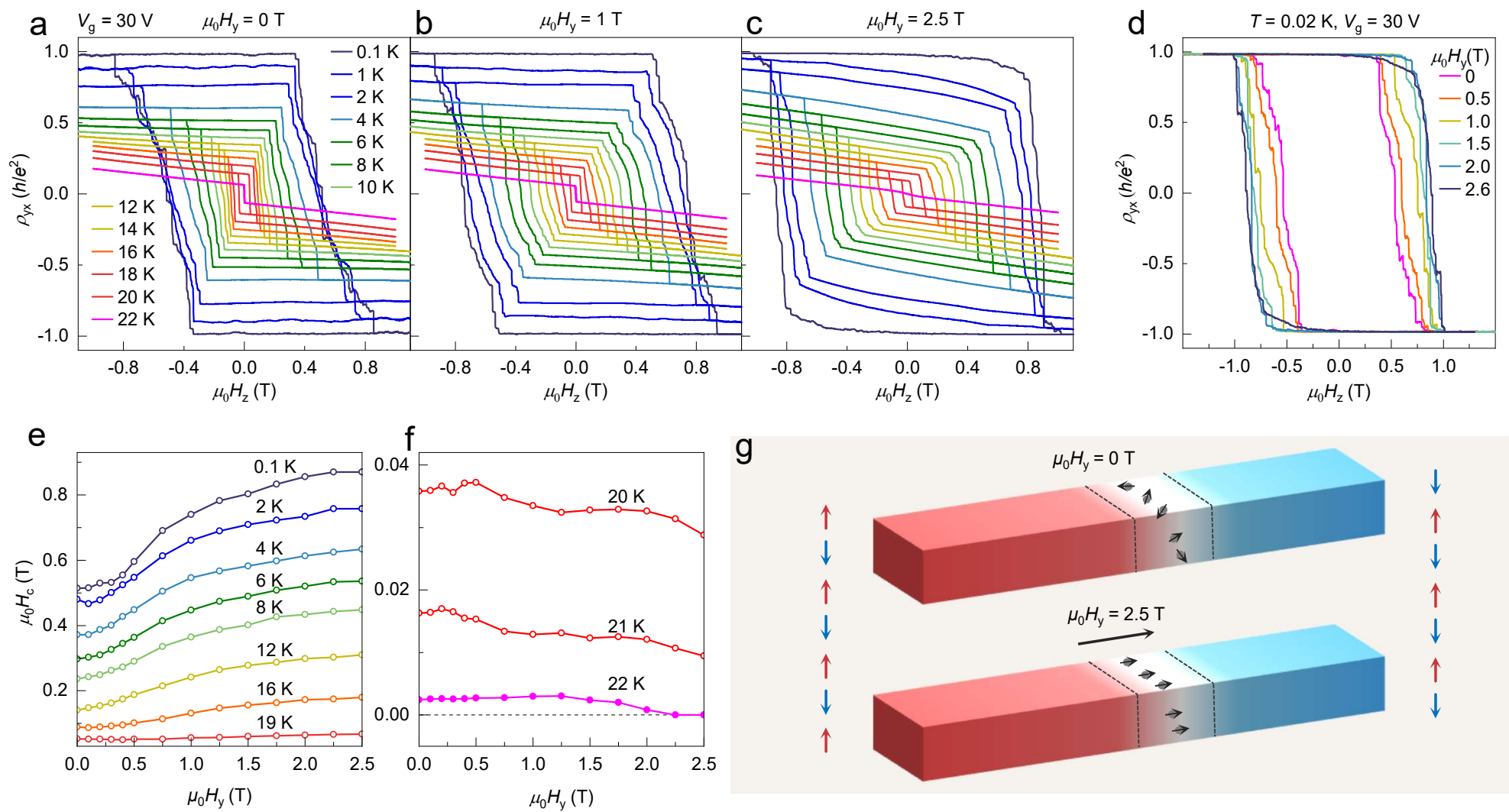

Fig. 3

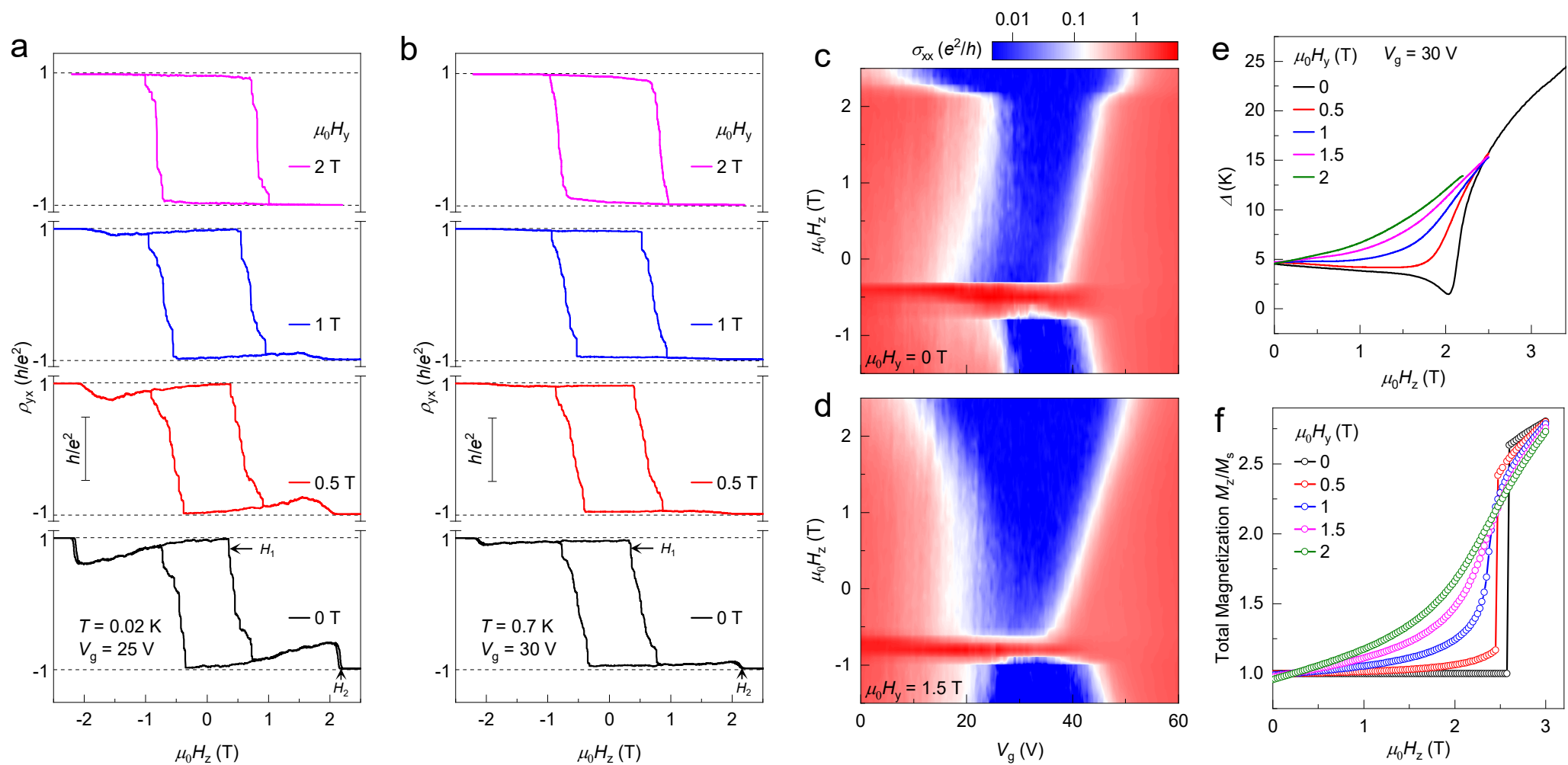

Fig. 4